\documentclass[english,aps,twocolumn,superscriptaddress]{revtex4}
\pdfoutput=1
\usepackage{lmodern}

\usepackage[T1]{fontenc}
\usepackage[latin9]{inputenc}
\usepackage{textcomp}
\usepackage{pmboxdraw}
\usepackage{amsthm}
\usepackage{amsmath}
\usepackage{graphicx}
\usepackage{esint}

\makeatletter

\newcommand{\lyxmathsym}[1]{\ifmmode\begingroup\def\b@ld{bold}
  \text{\ifx\math@version\b@ld\bfseries\fi#1}\endgroup\else#1\fi}

\@ifundefined{textcolor}{}
{%
 \definecolor{BLACK}{gray}{0}
 \definecolor{WHITE}{gray}{1}
 \definecolor{RED}{rgb}{1,0,0}
 \definecolor{GREEN}{rgb}{0,1,0}
 \definecolor{BLUE}{rgb}{0,0,1}
 \definecolor{CYAN}{cmyk}{1,0,0,0}
 \definecolor{MAGENTA}{cmyk}{0,1,0,0}
 \definecolor{YELLOW}{cmyk}{0,0,1,0}
 }

\pdfoutput=1

\makeatother

\usepackage{babel}

\begin{document}

\preprint{This line only printed with preprint option}

\title{An entangled-LED driven quantum relay over 1 km}

\author{C. Varnava}

\address{Toshiba Research Europe Limited, 208 Science Park, Milton Road, Cambridge
CB4 0GZ, UK}

\address{Cambridge University Engineering Department, 9 J J Thomson Avenue,
Cambridge CB3 0FA, UK}

\author{R. M. Stevenson}

\email{mark.stevenson@crl.toshiba.co.uk}

\address{Toshiba Research Europe Limited, 208 Science Park, Milton Road, Cambridge
CB4 0GZ, UK}

\author{J. Nilsson}

\author{J. Skiba-Szymanska}

\author{B. Dzur\v{n}\'{a}k}

\altaffiliation{Current Address: Department of Physics and Astronomy, University of Sheffield, Sheffield S3 7RH, UK}

\author{M. Lucamarini}

\address{Toshiba Research Europe Limited, 208 Science Park, Milton Road, Cambridge
CB4 0GZ, UK}

\author{R. V. Penty}

\address{Cambridge University Engineering Department, 9 J J Thomson Avenue,
Cambridge CB3 0FA, UK}

\author{I. Farrer}

\author{D. A. Ritchie}

\address{Cavendish Laboratory, University of Cambridge, J J Thomson Avenue,
\\
Cambridge CB3 0HE, UK}

\author{A. J. Shields}

\address{Toshiba Research Europe Limited, 208 Science Park, Milton Road, Cambridge
CB4 0GZ, UK}
\begin{abstract}
Quantum cryptography allows confidential information to be communicated
between two parties, with secrecy guaranteed by the laws of nature
alone. However, upholding guaranteed secrecy over quantum communication
networks poses a further challenge, as classical receive-and-resend
routing nodes can only be used conditional of trust by the communicating
parties. Here, we demonstrate the operation of a quantum relay over
1 km of optical fiber, which teleports a sequence of photonic quantum
bits to a receiver by utilizing entangled photons emitted by a semiconductor
LED. The average relay fidelity of the link is 0.90\textpm{}0.03,
exceeding the classical bound of 0.75 for the set of states used,
and sufficiently high to allow error correction. The fundamentally
low multi-photon emission statistics and the integration potential
of the source present an appealing platform for future quantum networks.
\end{abstract}
\maketitle
The critical importance of information security in the digital age
has led to the pervasive use of cryptography. The emerging field of
quantum cryptography\citep{Gisin2002,Scarani2009}, offers a means
to guarantee the security of digital interactions which can be proven
information theoretic secure. Reported quantum key distribution systems
are typically based on weak-coherent optical pulses, and have evolved
rapidly. Such systems allow unique cryptographic keys to be shared
between directly connected users on point-to-point\citep{Rosenberg2007,Dixon2010,Wang2012}
or point-to-multipoint links\citep{Frohlich2013}. 

In contrast, the situation for networks connecting multiple parties
is less developed. One solution has been to demonstrate nodal networks,
based on individual quantum links connected by classical intermediary
systems\citep{Peev2009}. The result of this type of network topology
is that end users are required to trust the intermediary systems,
which arguably diminishes the value of the underlying quantum cryptography.

To establish fully quantum multi-partite networks, it is necessary
to route quantum signals through a backbone of quantum nodes. This
can be achieved by leveraging quantum entanglement to set up non-local
correlations between measurements by end users. Examples of such schemes
are distribution of entangled photon pairs to end users, where local
measurements are performed\citep{Ekert1991}, or conversely, where
photons are sent by two users to be projected into a Bell state by
an intermediate quantum node\citep{Braunstein2012,Lo2012,Tang2014}.
Photonic quantum repeaters\citep{briegel1998} and relays\citep{Jacobs2002}
employ both of these effects to teleport entangled or single qubits
respectively in a manner that can be chained to create a fully quantum
network for which theoretically proven quantum security can be preserved. 

\begin{figure}
\includegraphics[width=8.5cm]{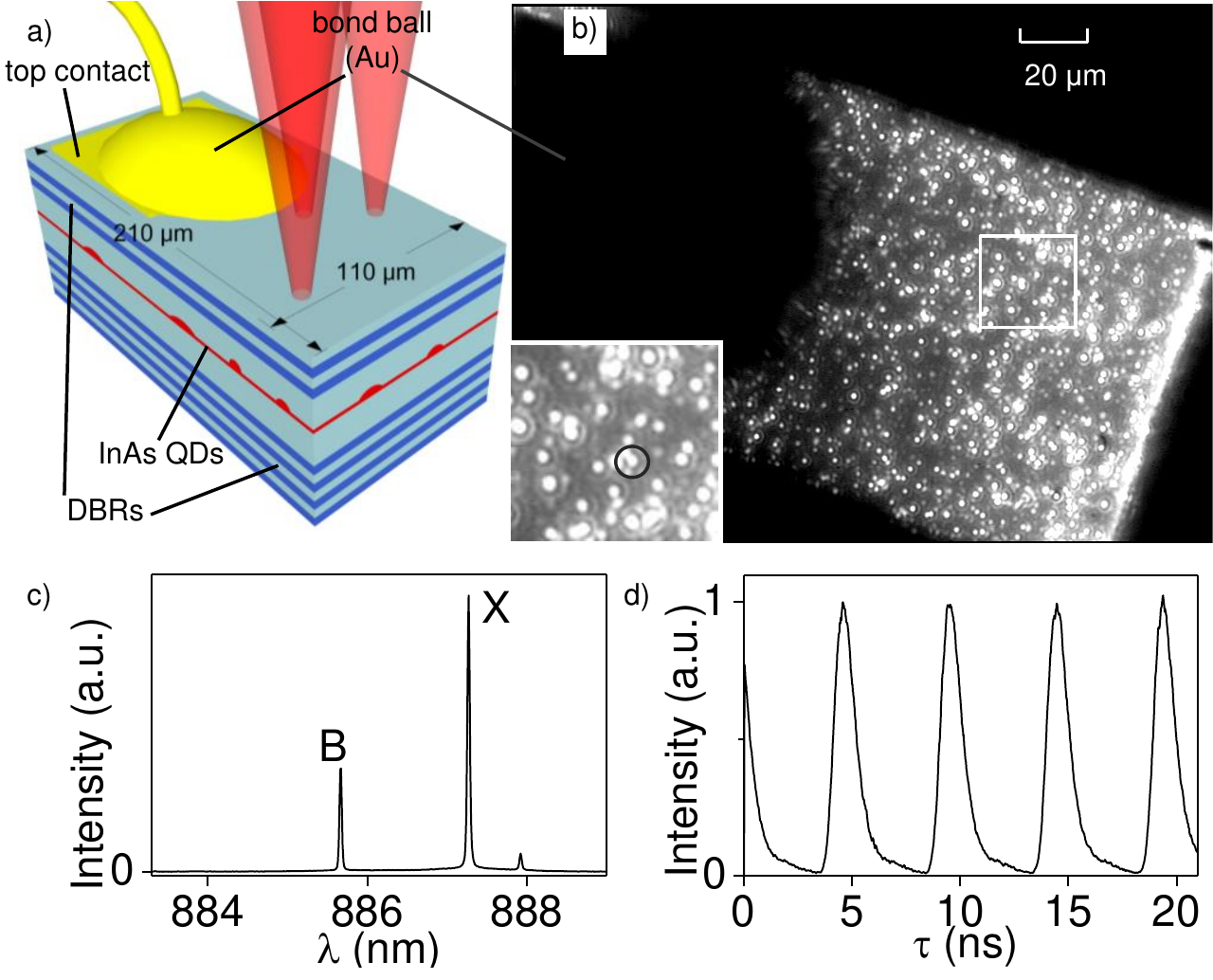}\caption{(a) Schematic of the ELED used in this experiment (not to scale).
The entangled photon source is an InAs quantum dot (QD) embedded in
a p-i-n diode structure with a cavity centred at \textasciitilde{}886nm,
between top and bottom distributed Bragg reflectors (DBRs). (b) Microscope
image of the mesa used for this experiment. Individual quantum dots
can be identified as points of light in the image. The dark circular
area is the gold bond to the top contact. The quantum dot chosen for
the experiment is identified in the inset. (c) Biexciton (B) and exciton
(X) electroluminescence spectrum under experimental biasing conditions.
(d) Time-resolved biexciton electroluminescence.}

\end{figure}

Here, we report operation of a quantum relay over 1 km of optical
fiber using entangled photons generated by a light emitting diode
to teleport photonic qubits encoded on weak coherent pulses emitted
by a laser. Compared to previously reported quantum relays\citep{DeRiedmatten2004}
and photonic teleportation over significant distances\citep{Ma2012,Bussieres2014},
our system is directly electrically driven using a simple semiconductor
device, offering a route to large-scale network deployments. Teleporting
weak coherent states offers potential enhancements to state-of-the
art quantum key distribution systems, as it creates output photons
with sub-Poissonian statistics immune to the photon number splitting
attack\citep{Brassard2000,Stevenson2013}, and protects against intrusions\citep{Lo1999}.

At the heart of our quantum relay is an entangled-light-emitting diode
(ELED)\citep{salter2010}, as shown in Fig. 1(a). It consists of a
layer of self-assembled indium arsenide quantum dots within a gallium
arsenide microcavity (Appendix). We have optimized the resistance
and capacitance of the device to allow it to be driven by short electrical
pulses, without compromising entanglement fidelity or photon coherence.
This has allowed electrically triggered quantum teleportation using
an LED, which has previously been limited to only d.c. operation\citep{Nilsson2013,Stevenson2013}.

An image of the ELED in operation is shown in Fig. 1(b). Individual
points of light are observed, corresponding to light emission from
individual quantum dots. We select emission from a chosen quantum
dot, indicated in the inset image, by collection with a single mode
fibre. The emission spectrum measured by a grating spectrometer and
CCD camera is shown in Fig. 1(c). Two strong emission lines are observed
corresponding to the first, biexciton photon ($B$) and second, exciton
photon ($X$) emission. The $B$ and $X$ emission lines are then
spectrally filtered with a diffraction grating to isolate them from
each other and other emission from the device, including that originating
from the quantum well-like wetting layer, and any other nearby quantum
dots (such as the weak peak seen at \textasciitilde{}888 nm).

The ELED was driven at a repetition rate of 203 MHz, with pulses of
nominally 0.4V amplitude and 490\,ps duration. Time-resolved electroluminescence
was measured under these conditions and shown in Fig. 1(d) for $B$
as a black line. The emission is strongly pulsed, and well contained
within each cycle.

The experimental quantum relay system is shown in Fig. 2(a). It comprises
4 sections separated by three 350\,m fiber optic links. The first,
\textquoteleft{}Sender\textquoteright{} section is thus separated
from the last \textquoteleft{}Receiver\textquoteright{} section by
1.05 km of optical fiber. Entangled photons emitted by the ELED are
divided, the B photons are sent to the Bell-State Measurement (BSM)
section, and the $X$ photons are sent to the \textquoteleft{}Receiver\textquoteright{}.
The \textquoteleft{}Sender\textquoteright{} employs a wavelength tunable
c.w. laser diode, from which pulses are generated by an external optical
intensity modulator, tuned to the frequency of the ELED. The polarized
pulses are then rotated by a polarization controller PC1 to encode
the qubit, before transmission to the \textquoteleft{}BSM\textquoteright{}
section.

\begin{figure}
\includegraphics[width=8.5cm]{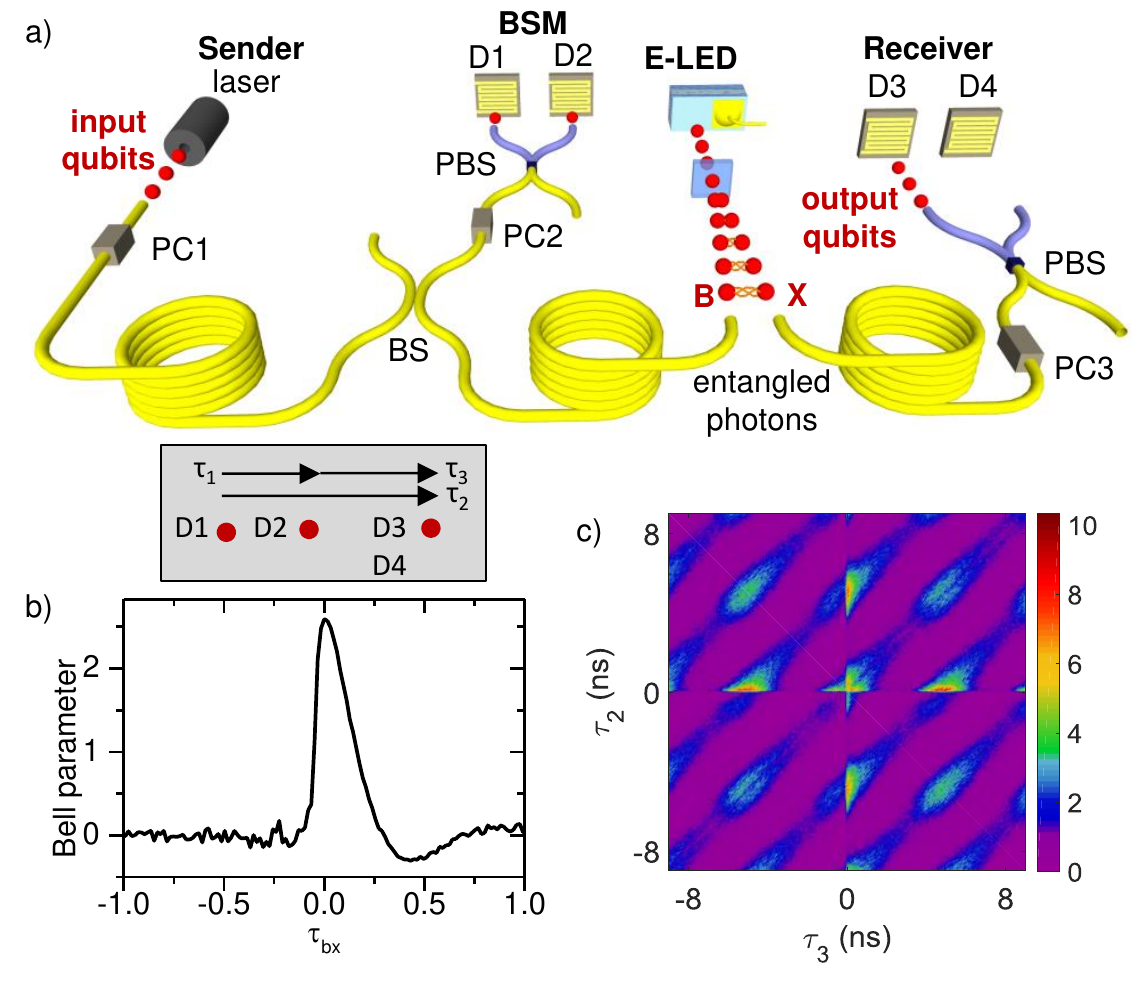}

\caption{(a) Quantum relay experimental setup. The Sender encodes photons from
an externally modulated laser diode with qubit states $|H\rangle$,$|V\rangle$,$|D\rangle$,$|A\rangle$,
using a polarization controller PC1, to be transferred to the Receiver
by quantum teleportation. Sender and Receiver are separated by \textasciitilde{}1\,km
of fiber and a Bell-State Measurement (BSM) node in between. The BSM
and Receiver share an entangled pair of photons emitted from the ELED.
The input qubits interfere with biexciton photons on beamsplitter
BS. Once detectors D1 and D2 measure their state, teleported output
qubits are detected with a polarizing beamsplitter (PBS) at D3 and
D4. All state calibrations at each node are done with polarization
controllers (PC). (b) Bell parameter extracted from experimental data.
(c) Average third-order correlation $g^{(3)}$. Pulsed character of
the correlations is observed. The single-photon property of emission
is seen as a dip in coincidences at $\tau_{2}=\tau_{3}$. Higher three-photon
coincidences running along $\tau_{2}=0$ and $\tau_{3}=0$ originate
from elevated probability of the ELED emitting a pair of photons simultaneously.}

\end{figure}

At the \textquoteleft{}BSM\textquoteright{} section, an imbalanced
beamsplitter BS combines 95\% of the $B$ photons with 5\% of the
laser photons into one output arm, before a polarization controller
PC2 and polarizing beamsplitter projects horizontal ($H$) and vertical
($V$) polarized photons onto superconducting single photon detectors
(SSPD) D1 and D2. The function of this section is to perform a Bell
state measurement in the state $(|H_{L}V_{B}\rangle+|V_{L}H_{B}\rangle)/\sqrt{2}$
where subscripts $L$ and $B$ denote photons originating from the
laser and biexciton respectively. Such a measurement collapses the
formerly entangled $X$ photon into a quantum state related to the
input qubit together with a trivial unitary transformation\citep{bennett1993}.
In this work, input qubits of the form $\cos(a)|H_{L}\rangle+e^{ib}\sin(a)|V_{L}\rangle$
are teleported to the state $\cos(a)|V_{X}\rangle+e^{ib}\sin(a)|H_{X}\rangle$.

Theoretical analysis of the intricate time-dependent three-photon
fields reveals a range of laser pulse conditions for which the calculated
peak teleportation fidelities are close to optimum (Appendix). The
laser pulse properties observed at detectors D1 and D2 were set accordingly,
with integrated intensity of 0.85 relative to the $B$ pulse, a fitted
Gaussian FWHM of 0.95\textpm{}0.01\,ns, and a delay of the laser
pulse after the $B$ pulse of 0.60\textpm{}0.02\,ns relative to maximum
overlap.

The entanglement properties of the photons emitted by the ELED were
characterized within the quantum relay system. Second-order correlation
measurements were performed in the rectilinear \{$H$,$V$\} and diagonal
\{$D$,$A$\} linear polarization bases, where $D$ and $A$ represent
the diagonal and anti-diagonal polarizations respectively. From these
measurements it is possible to determine Bell\textquoteright{}s parameter
to indicate the degree of entanglement present\citep{Clauser1969,young2009}.
The result is shown in Fig. 2(b) as a function of the time delay $\tau_{BX}$
between a biexciton photon detected at D1 or D2, and an exciton photon
at D3 or D4. For simultaneously detected photons, a Bell parameter
of 2.59\textpm{}0.01 is observed, exceeding the limit of 2 for classical
behavior, and corresponding to 91.8 \% of the ideal value of $2\sqrt{2}$.
To our knowledge, this is the first time entanglement has been distributed
by a quantum dot source over a distance longer than a few meters.

The quantum relay was operated using polarization encoded BB84 quantum
states\citep{bennett1984}, using the rectilinear and diagonal bases.
Three-photon detection statistics were recorded between a pair of
photons at D1 and D2, and one at D3 or D4. The polarizations of the
input qubit and measurement basis, controlled by PC1 and PC3 respectively,
were switched during the experiment so that a randomized sequence
of teleported qubits could be recorded.

Fig. 2(c) shows the measured third-order correlation function $g^{(3)}$
averaged over all 4 polarization inputs, and over the corresponding
co- and cross- polarized outputs. The horizontal and vertical axes
are the time delays $\tau_{2}$ and $\tau_{3}$ between photon detection
at D1 or D2 respectively, and a photon at D3 or D4. The intensity
distribution has highly pulsed character, with peaks occurring when
the delays $\tau_{2}$ and $\tau_{3}$ are an integer multiple of
the \textasciitilde{} 4.9\,ns repetition period. This is in contrast
to previous reports of quantum teleportation with quantum dots, which
operated in continuous mode. Stronger intensity peaks are observed
for $\tau_{2}=0$ and $\tau_{3}=0$, as exciton emission directly
following biexciton emission is enhanced. For coincident detection
at D1 and D2, $\tau_{2}=\tau_{3}$, a low intensity line indicates
a suppression in detecting two $B$ photons simultaneously due to
the sub-Poissonian nature of the ELED source.

\begin{figure}
\includegraphics[width=5.5cm]{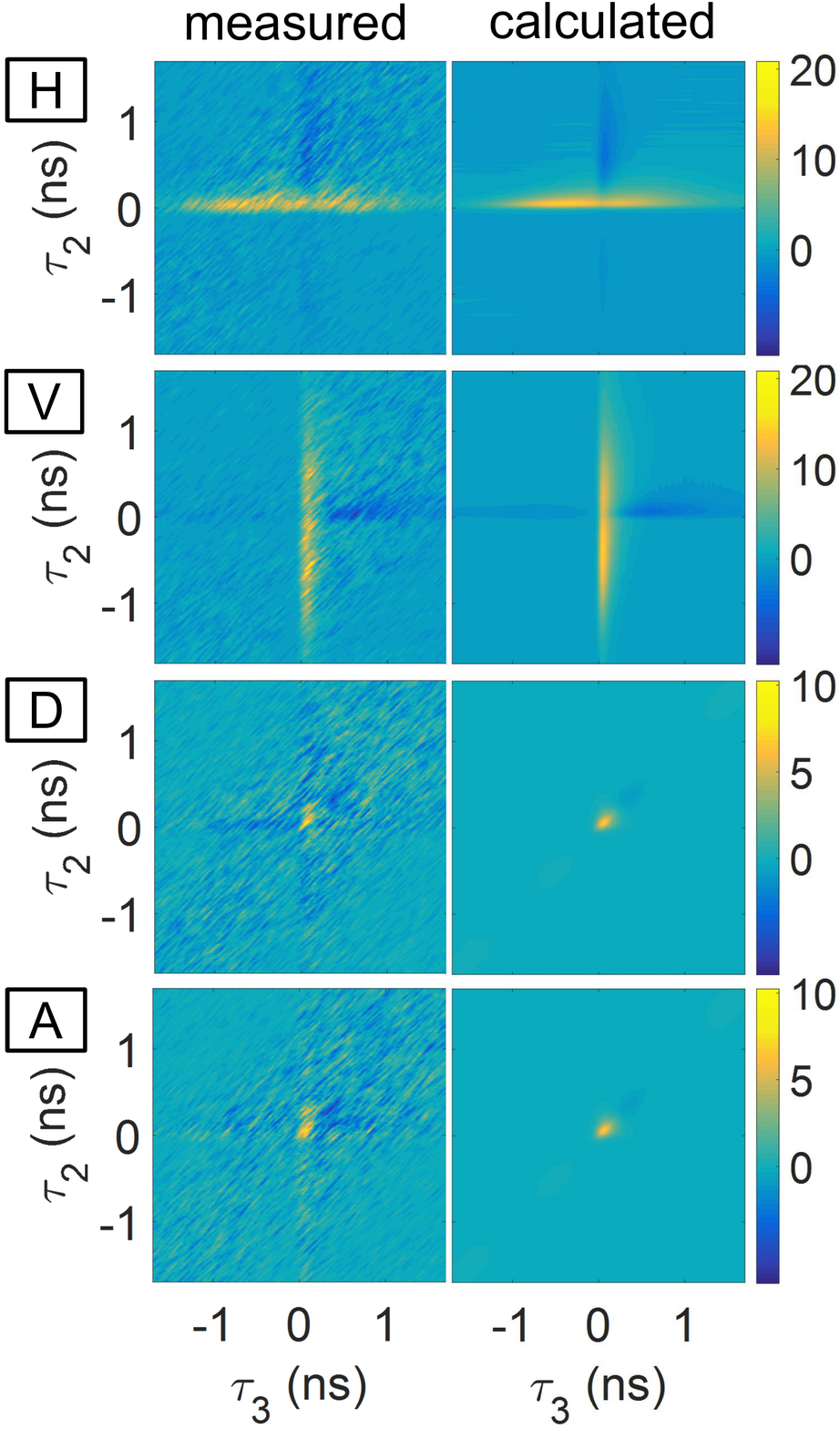}

\caption{Difference between expected and unexpected third-order correlations
$g^{(3)}$ as a function of photon detection time delays $\tau_{2}$
and $\tau_{3}$. Measured and calculated results are shown in the
left and right column respectively, corresponding to input states
H, V, D, and A (from top to bottom).}

\end{figure}

The polarization dependence of the quantum relay is evaluated from
the difference between third-order correlation measurements with the
expected, and unexpected (i.e. orthogonal to expected) output polarizations.
Such $g^{(3)}$ contrast measurements are presented in Fig. 3 as a
function of $\tau_{2}$ and $\tau_{3}$ for each of the input qubit
states H,V,D, and A. High contrast at $\tau_{2}=0$ ($\tau_{3}=0$)
is observed due to H (V) input laser photons exciting the D1 (D2)
detector, so that contrast is dominated by correlation between $X$
and a $B$ photon at D2 (D1). The correlation contrast for D and A
however looks quite different, dominated by a peak centered at $\tau_{3}=\tau_{2}=0$,
as two-photon interference between simultaneously detected photons
is required for teleporting states in a superposition state. Calculations,
shown in the right column, agree well with observations (Appendix).

The performance of the quantum relay is assessed by determining the
relay fidelity $F_{P}$, which for each input photon state $P$ is
determined by:

\[
F_{P}(\tau_{3},\tau_{2})=g_{P'}^{(3)}(\tau_{3},\tau_{2})/(g_{P'}^{(3)}(\tau_{3},\tau_{2})+g_{Q'}^{(3)}(\tau_{3},\tau_{2})),\]
where $P'$ and $Q'$ are the expected and orthogonal unexpected output
polarizations respectively. Averaging across the four input states
H,V,D, and A gives the average relay fidelity $F$, which is plotted
in Fig. 4(a). A clear peak is observed around $\tau_{2}=\tau_{3}=0$,
where the fidelity clearly exceeds 0.75, the limit for optimal classical
teleportation schemes using four-state protocols. Calculated behavior,
shown in Fig. 4(b), shows similar features. The maximum measured value
of $F$ is more clearly observed in Fig. 4(c), which plots $F$ as
a function of $\tau_{2}$ or $\tau_{3}$ for simultaneous detection
of two photons at the \textquoteleft{}BSM\textquoteright{} section.
The peak rises to a maximum value of 0.900\textpm{}0.028, exceeding
the six-state average fidelity reported previously\citep{Stevenson2013}.
The corresponding measured individual relay fidelities are 0.957\textpm{}0.042
and 0.951\textpm{}0.0475 from polar states H and V, and 0.845\textpm{}0.064,
0.847\textpm{}0.063 for superposition states D and A.

\begin{figure}
\includegraphics[width=5.5cm]{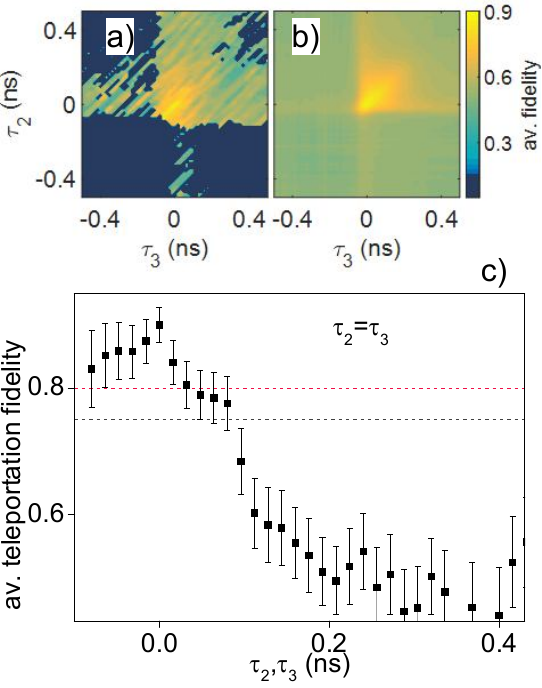}

\caption{(a,b) Relay fidelity averaged across four input qubit states H, V,
D, and A, as a function of the time delays $\tau_{2}$ and $\tau_{3}$.
(a) and (b) shown experimental and simulated results respectively.
Note the measured fidelity is not defined some regions of the plot
as no photons were detected due to the ELED or laser being \textquoteleft{}off\textquoteright{}
(dark blue). (c) Average relay fidelity as a function of time delay
$\tau_{2}$ and $\tau_{3}$ for $\tau_{2}=\tau_{3}$. High fidelity
points are concentrated around $\tau_{2}$,$\tau_{3}$=0, where the
measured fidelity exceeds the classical limit of 0.75 shown by black
dashed line. The threshold for 4-state error correction of 0.8 is
shown as dashed red line. }

\end{figure}

The high measured teleportation fidelities confirm that Sender and
Receiver have shared information in excess of any information held
by an eavesdropper. The difference between unity and the measured
fidelity is the quantum bit error rate (QBER) of the shared key. Error
correction protocols are known that can correct for QBER up to 0.2
in 4-state protocols\citep{Chau2002}, which corresponds to a minimum
relay fidelity of 0.8, well below the experimentally measured value.

For smaller QBER, as in our experiment, more effective error correction
protocols are available\citep{Brassard1994}. Using the single-photon,
efficient BB84 protocol\citep{Lo2005}, in the limit of infinitely
many signals shared by the users, the fraction $R$ of secure bits
that can be extracted from each detected photon is\citep{Koashi2009}:
\[
R=1-h(Q_{Z})-h(Q_{X}),\]
where $h$ is the binary entropy function. By replacing the QBERs
$Q_{Z}$ and $Q_{X}$ in the $Z$ and $X$ basis by those measured
of 4.6\% and 15.4\% in the \{$H$,$V$\} and \{$D$,$A$\} bases respectively,
we find that 0.111 secure key bits can be distilled from our Relay
per detected photon.

A key advantage of quantum-dot based entangled light sources over
spontaneous parametric down conversion is that they can in principle
operate \textquoteleft{}on-demand\textquoteright{} and deterministically
deliver a single entangled photon pair when triggered, without detrimental
additional pairs. In our experiments, we measure the second-order
correlation for coincident $X$ photons to be close to optimal at
$0.046\pm0.008$, highly suppressed from the Poissonian value of 1.
Achieving maximal efficiency is predominantly limited by the photon
collection efficiency, and the temporal post-selection window that
heralds the occurrence of a teleportation event. Very high photon
collection efficiencies have been reported in optically driven quantum
dot nanostructures\citep{dousse2010,Claudon2010}, for which electrical
injection schemes could be developed. Increasing the temporal region
of high fidelity may be achieved by matching the laser and biexciton
pulse shapes, reducing the excitation pulse width, reducing the fine-structure-splitting
of 4.2\textpm{}0.1\,\textmu{}eV, and extending the coherence time
of the biexciton photons, which at 141.6\textpm{}4.2\,ps limits the
fidelity of teleporting D and A states.

In conclusion, we have reported the operation of a 203\,MHz clocked
quantum relay over 1 km of optical fiber using an electrically driven
semiconductor light source. The measured relay fidelity of laser generated
photonic qubits of 0.900\textpm{}0.027 exceeds both the threshold
for quantum behavior, and the one for generating error corrected quantum
keys with the BB84 protocol. Coupled with advances of quantum dot
based entangled light sources at telecom wavelengths, our results
suggest ELED technology could meet the need for a practical solution
to create the backbone of future quantum networks.
\begin{acknowledgments}
The authors would like to thank M. Razavi for theoretical support
and C. Panayi for useful discussions. The authors would like to acknowledge
partial financial support through the UK EPSRC and the EU Marie Curie
Initial Training Network Spin-optronics. 
\end{acknowledgments}
\bibliographystyle{apsrev}
\bibliography{1kmrelay}

\section*{Appendix}

\subsection*{Entangled light source}

Entangled-photon pairs are generated by an InAs quantum dot, embedded
in a 2-lambda GaAs optical cavity within a p-i-n heterostructure grown
by MBE. In order to improve light collection, 6 top and 18 Bragg reflectors
(DBRs) were grown. The relatively small dimensions of the mesa (210x110\,\textmu{}m)
allow for a simple direct bonding design, while increasing the pulsed
operation performance compared to previously reported, larger, entangled-LEDs.
The dot density is low enough so that no apertures were required and
the light was collected using a single-mode fiber. The ELED device
was driven by electrical injection at a frequency of 203\,MHz, in
forward-bias, with a modulating a.c. voltage of nominally square pulses
with 0.4 V amplitude and 490 ps width, at a temperature of 19.7\,K.
The emission spectrum of the dot under these conditions shows the
biexciton $B$ and exciton $X$ at 885.7 and 887.3\,nm respectively.
The fine-structure-splitting was 4.2\textpm{}0.1\,\textmu{}eV.

\subsection*{Quantum relay}

For the input qubit photons at the Sender section, a CW laser was
externally modulated and synchronized with the dot driving frequency
using a Mach-Zehnder optical intensity modulator. The generated pulses
can have a FWHM of 0.60\textpm{}0.02\,ns. The relative time-integrated
intensities between laser and biexciton photons incident on detectors
D1 and D2 was set to 0.85:1. The input qubit polarization state was
selected using a pseudo-random number generator and polarization controller
PC1 at a frequency exceeding the 3-photon coincidence rate. We note
that the system is in principle compatible with quantum random number
generators and phase modulation at frequencies exceeding the input
qubit rate. The logical teleportation states \{$H,V$\} are calibrated
to the quantum dot\textquoteright{}s polarization eigenbasis, and
the diagonal states \{D,A\} are set at +/-45� to the rectilinear basis
using a linear polarizer.

A 95:5 fiber beamsplitter was used to interfere the input laser photons
with the biexciton photons. We only look at events from one output
arm of this beamsplitter which maximizes the fraction of ELED photons
detected in our experiment. The system was actively stabilized using
electrical polarization controllers at each of the sections during
the experiment. The degree of polarization for the input control qubits
was maintained at 97.5\textpm{}1.5\%. The 5\% output port was used
to spectrally tune the laser to the biexciton wavelength using a grating
spectrometer, with an average detuning of 0.12\textpm{}1.68\,\textmu{}eV
throughout the experiment.

Finally, photons at the BSM and Receiver polarizing beamsplitters
(PBSs) were detected using four superconducting detectors (SSPDs).
The timing jitter between pairs of detectors was approximately Gaussian
with FWHM of 58.4\textpm{}2.5\,ps, with single photon counting hardware
resolution of 16ps.

\subsection*{Three-photon correlations}

Three-photon coincidences were recorded corresponding to 2 photons
at BSM detectors D1 and D2, with polarization H and V, with one photon
at either Receiver\textquoteright{}s detectors D3 and D4. D3 and D4
record both expected photons with polarization P\textquoteright{}
and unexpected output photons with polarization Q\textquoteright{}
simultaneously.

The detection times at each detector are defined as $t_{1}$ (D1),
$t_{2}$ (D2), and $t_{3}$ (D3 \& D4). All events were recorded relative
to detections at D1 (H polarized photons). Third-order correlations
$g^{(3)}$ for each output polarization were determined from the normalized
statistics of the three photon coincidences, as a function of two
time delays, $\tau_{2}=t_{3}-t_{1}$ and $\tau_{3}=t_{3}-t_{2}$.
We note that the choice of which pair of time delays amongst 3 photons
to choose is somewhat arbitrary, for example $t_{2}-t_{1}$ and $t_{3}-t_{1}$
have been used in previous reports.

\subsection*{Error analysis}

Errors are dominated by Poissonian counting statistics, which determines
the errors on the third-order correlations from the number of photons
detected. Errors are propagated to determine the relay fidelity errors
of individual input states, and the average relay fidelity. Systematic
errors due to the temporal calibration and resolution of our system
are also included in the fidelity measurements, but are almost negligible.
Note that all experimental 3-photon results are presented for a time-integration
window of 32\texttimes{}112\,ps in $\tau_{1}$ and $\tau_{2}$ respectively,
and evaluated on a 16\,ps grid, so adjacent points are not independent.

\subsection*{Calculation of time-dependent $g^{(3)}$ correlations}

In our experiments, we employ a beamsplitter to combine laser photons
with biexciton photons originating from an entangled-LED, as shown
in figure 5(a). To conserve biexciton photons from the entangled light
source, the beamsplitter is highly imbalanced such that the transmission
coefficient $|k_{t}|\gg|k_{r}|$. Only photons emerging on the efficient
arm of the beamsplitter are measured, so for simplicity, the transmission
and reflection coefficients of the beamsplitter, and phase change
due to the coupling, are included within the amplitude and overall
phase components of the laser and quantum dot wavefunction $\psi_{L}$
and $\psi_{BX}$.

\begin{figure}[b]
\includegraphics[width=8.5cm]{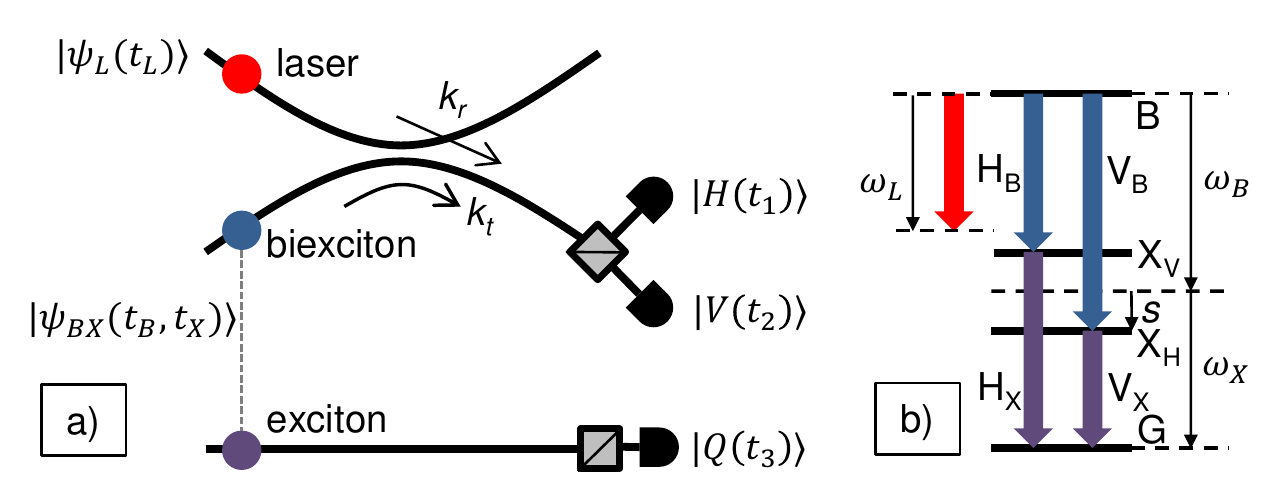}

\caption{(a) Schematic of quantum relay scheme. Laser and biexciton photons
meet at a fiber optic coupler with transmission and coupling amplitudes
$k_{t}$ and $k_{r}$, before direction by a polarizing beamsplitter
to a pair of photon detectors. The exciton photon polarization is
selected in a given state $Q$, and measured by a third detector.
(b) Level diagram showing the biexciton $(B)$, exciton $(X)$ and
ground $(G)$ states of a single quantum dot, plus the input laser
photon. Frequencies and detunings of various photon states are shown.}

\end{figure}

To begin, we define the laser and quantum dot states. In general,
they are given by: \[
\psi_{j}=A_{j}(t_{j})C_{j}(t_{j})|\Psi_{j}(t_{j})\rangle,\]
where $A_{j}(t_{j})$ is the time dependent real amplitude, $C_{j}(t_{j})$
is the overall phase, and $|\Psi_{j}(t_{j})\rangle$ the polarisation
and any polarisation dependent phase. The overall phase term is further
defined as:

\[
C_{j}(t_{j})=e^{i\omega_{j}t_{j}}e^{i\phi_{j}(t_{j})}.\]
The first exponent is coherent, and contributes to the final solutions
only through detuning of $\omega_{j}$ to the primary frequencies
$\omega_{B}$ and $\omega_{X}$. Decoherence is represented by random
fluctuations of the phase $\phi_{j}$ as a function of time $t_{j}$,
such that $\langle e^{i[\phi_{j}(t_{j})-\phi_{j}(t_{j}+\Delta)]}\rangle=e^{-|\Delta|/T_{j}}$,
where $T_{j}$ is the coherence time of photon $j$\citep{Legero2003,Santori2004}.

For the laser input state with polarization defined by real parameters
$a$ and $b$, we have: \begin{gather*}
\psi_{L}=A_{L}(t_{L})C_{L}(t_{L})|\Psi_{L}\rangle,\\
|\Psi_{L}\rangle=cos(a)|H\rangle+e^{ib}sin(a)|V\rangle.\end{gather*}

The output polarization state is:\[
|\Psi_{Q}\rangle=cos(x)|H\rangle+e^{iy}sin(x)|V\rangle.\]

For photon pairs from the quantum dot, we have:\[
\psi_{BX}=A_{BX}(t_{B},t_{X})C_{B}(t_{B})C_{X}(t_{X})|\Psi_{BX}(t_{B},t_{X})\rangle.\]

The ideal quantum dot biphoton amplitude and state are given by\begin{gather*}
A_{BXe}(t_{B},t_{X}),\\
|\Psi_{BXe}\rangle=(e^{is(t_{X}-t_{B})}|HH\rangle+e^{-is(t_{X}-t_{B})}|VV\rangle)/\sqrt{2}.\end{gather*}

In practice, however, the emission from the quantum dot is partially
mixed. We approximate this with the amplitude $A_{BXu}(t_{B},t_{X})$
and equal mixture of the polarization states:\begin{alignat*}{1}
\text{\ensuremath{\Psi}}_{HH} & =e^{is(t_{X}-t_{B})}|HH\rangle,\\
\text{\ensuremath{\Psi}}_{HV} & =e^{is(-t_{X}-t_{B})}|HV\rangle,\\
\text{\ensuremath{\Psi}}_{VH} & =e^{is(t_{X}+t_{B})}|VH\rangle,\\
\text{\ensuremath{\Psi}}_{VV} & =e^{is(-t_{X}+t_{B})}|VV\rangle.\end{alignat*}

The joint amplitude of the electric field at the three photon detectors
$Z$ is given by the following equation, where four-photon contributions
and higher are disregarded due to their relatively small probability
compared to three-photon events:\begin{multline*}
Z(t_{1},t_{2},t_{3})=\langle H\lyxmathsym{\textSFxi}\psi_{L}(t_{1})\langle VQ\lyxmathsym{\textSFxi}\psi_{BX}(t_{2},t_{3})\rangle\\
+\langle V\lyxmathsym{\textSFxi}\psi_{L}(t_{2})\rangle\langle HQ\lyxmathsym{\textSFxi}\psi_{BX}(t_{1},t_{3})\rangle\\
+\langle HV\lyxmathsym{\textSFxi}\psi_{LL}(t_{1},t_{2})\rangle\langle Q\lyxmathsym{\textSFxi}\psi_{X}(t_{3})\rangle\\
+\langle HVQ|\psi_{BBX}(t_{1},t_{2},t_{3})\rangle.\end{multline*}
The first two terms are the desired three-photon amplitudes originating
from a single laser, biexciton, and exciton photon. The third and
fourth terms originate from two laser photons plus one exciton photon,
and two biexciton photons plus one exciton photon respectively.

The three-photon intensity $Z(t_{1},t_{2},t_{3})Z^{*}(t_{1},t_{2},t_{3})$
is evaluated and integrated over the arrival time of the $X$ photon
$t_{3}$, from which we drop the subscript for convenience. Finally,
we make substitutions of the form: \begin{align*}
\eta_{j}I_{j}(t_{j}) & =A_{j}^{2}(t_{j}),\\
\eta_{j}\eta_{k}g_{jk}^{(2)}(t_{j},t_{k}) & =A_{jk}^{2}(t_{j},t_{k}),\\
\text{\ensuremath{\eta}}_{j}\eta_{k}\text{\ensuremath{\eta}}_{l}g_{jkl}^{(3)}(t_{j},t_{k},t_{l}) & =A_{jkl}^{2}(t_{j},t_{k},t_{l}),\end{align*}
where $\eta_{j}$ is the time averaged intensity of photon $j$, and
$I_{j}(t_{j})$ the normalised intensity of photon $j$ as a function
of time. The final expression for the third-order correlation is,\begin{multline*}
g^{(3)}(\tau_{2},\tau_{3})\propto\\
\frac{1}{2}cos^{2}(a)sin^{2}(x)\int_{0}^{p}I_{L}(t-\tau{}_{2})g_{BXe}^{(2)}(t-\tau_{3},t)dt\\
+\frac{1}{2}sin^{2}(a)cos^{2}(x)\int_{0}^{p}I_{L}(t-\tau_{3})g_{BXe}^{(2)}(t-\tau_{2},t)dt\\
+\frac{1}{4}sin(2a)sin(2x)e^{-\frac{|\tau_{1}|}{T_{L}}-\frac{|\tau_{1}|}{T_{B}}}\\
\times cos((\omega_{B}-\omega_{L})\tau_{1}-s(\tau_{3}+\tau_{2})+y+b)\\
\times\int_{0}^{p}\sqrt{(I_{L}(t-\tau_{2})I_{L}(t-\tau_{3})g_{BXe}^{(2)}(t-\tau_{2},t)g_{BXe}^{(2)}(t-\tau_{3},t)}dt\\
+\frac{1}{4}cos^{2}(a)\int_{0}^{p}I_{L}(t-\tau_{2})g_{BXu}^{(2)}(t-\tau_{3},t)dt\\
+\frac{1}{4}sin^{2}(a)\int_{0}^{p}I_{L}(t-\tau_{3})g_{BXu}^{(2)}(t-\tau_{2},t)dt\\
+\frac{\eta_{L}}{\eta_{B}}\frac{1}{4}cos^{2}(a)sin^{2}(a)\int_{0}^{p}g_{LL}^{(2)}(t-\tau_{2},t-\tau_{3})I_{X}(t)dt\\
+\frac{\eta_{B}}{\eta_{L}}\int_{0}^{p}g_{BBX}^{(3)}(t-\tau_{2},t-\tau_{3},t)dt.\end{multline*}

Note the final term containing $g_{BBX}^{(3)}(t-\tau_{2},t-\tau_{3},t)$
is evaluated by substitution with chains of two-photon correlations,
for different ordering of the two biexciton and one exciton photons.
This is justified as detection of the first photon places the system
to a well-defined state, which serves as the starting point for a
correlation to a second photon, which after detection again places
the system into another well-defined state, which is the starting
point with correlation to a third photon. A factor of 1/2 in the penultimate
term accounts for the Poissonian statistics of the laser two-photon
intensity.

\subsection*{Optimum teleportation conditions}

The single laser photon envelope $I_{L}$ was approximated as a Gaussian
function, which is a satisfactory approximation of what we observe
in experiment. Single biexciton and exciton photon intensities $I_{B}(t_{B})$
and $I_{X}(t_{X})$ were directly measured, and the fine-structure-splitting
and biexciton coherence time determined from polarization dependent
spectroscopy and single-photon interferograms respectively. Correlations
were measured between biexciton and exciton photons to determine ${g_{BX}^{(3)}(t_{B},t_{X}-t_{B})}$
for co- and cross-linearly polarized states. The entangled and unentangled
biphoton fractions $g_{BXe}^{(2)}$ and $g_{BXu}^{(2)}$ were extracted
from half the difference between, and uncorrelated component of, these
measurements respectively. Similarly correlations between pairs of
biexciton photons $g_{BB}^{(2)}(t_{1},t_{2})$ were directly measured.

Note that imperfections in polarization recovery and timing jitter
are well represented in the parameter data, as the same physical measurement
system was used for their measurement as for teleportation. However,
additional jitter was added when evaluating the calculations to terms
with discontinuities around zero delay. This fact, together with the
lower time-averaged estimate of the coherence time compared to biexciton
photons emitted in a single cycle, means that calculations are expected
to underestimate the relay fidelity slightly, as observed.

In order to maximize the relay fidelity in experiments, the expected
maximal fidelity was calculated as a function of laser pulse intensity,
width, and delay relative to the biexciton photon. The results are
summarized in figure 6, which plots the calculated relay fidelity
as a function of the pulse delay and width (a), with corresponding
optimal laser intensity in (b). Highest fidelities are observed for
a laser delayed between -0.6 and 0.8\,ns relative to maximal overlap
with the biexciton state, and for a pulse width between 0.45 and 2.05\,ns.
The values employed in our experiments are within this range.

\begin{figure}[b]
\includegraphics[width=8.5cm]{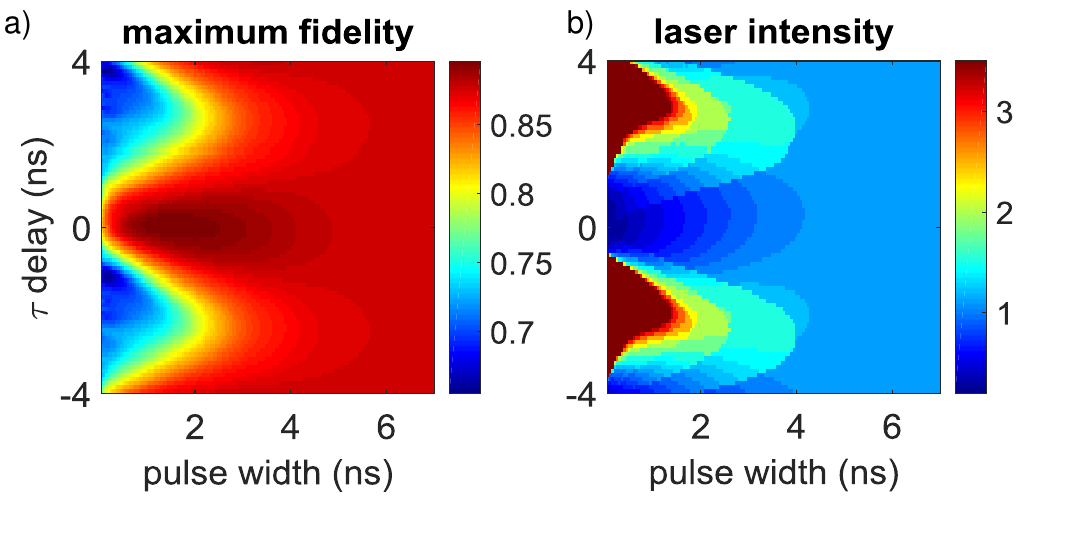}\caption{(a) Calculated maximum relay fidelity for our ELED with Gaussian laser
pulses as a function of laser pulse width and delay relative to the
biexciton photon. (b) Corresponding laser relative to biexciton intensity
required (up to a limit of 3.5) to achieve maximal fidelity.}

\end{figure}

\end{document}